\documentclass[12pt]{article}
\title{Some definite logarithmic integrals from Euler sums, and other integration
results}
\author{Mark W. Coffey\\
Department of Physics\\
Colorado School of Mines\\
Golden, CO  80401\\
(Received $\mbox{~~~~~~~~~~~~~~~~~~~~~~~~~~~~~~~2009}$)}
\date{December 28, 2009}
\begin{document}
\maketitle

\baselineskip=25pt

\begin{abstract}

We present explicit expressions for multi-fold logarithmic integrals that are
equivalent to sums over polygamma functions at integer argument.  Such relations
find application in perturbative quantum field theory, quantum chemistry, analytic
number theory, and elsewhere.  The analysis includes the use of properties of a
variety of special functions.

\end{abstract}

\centerline{\bf Key words and phrases}
\medskip

\baselineskip=15pt
\noindent
dilogarithm function, digamma function, polygamma function, harmonic numbers, Euler sums,
polylogarithm function, Hurwitz zeta function, Riemann zeta function, generalized
harmonic numbers, generalized hypergeometric function

\baselineskip=25pt
\bigskip
\centerline{\bf 2010 AMS classification numbers}
33B30, 33B15, 11M06 

\bigskip
\centerline{\bf 2010 PACS classification numbers}
02.30.Gp, 02.30.-f, 02.10.De

\pagebreak
\medskip
\medskip


\centerline{\bf Introduction}

It is well known that sums over polygamma functions at integer arguments, as well
as a vast array of logarithmic and polylogarithmic integrals, are useful for 
applications in quantum field theory and elsewhere \cite{coffey05,coffeyjmp2,duke}.
Additionally motivated by multiple-logarithm integrals occurring very recently in 
quantum chemistry \cite{bernie}, (see Propositions 2 and 6) we provide details for certain one-dimensional sums over polygamma functions.  Our treatment overlaps several areas of analysis, special function theory, analytic number theory \cite{apostol,matsuoka}, and integration \cite{ded}.  This is evidenced in the first Proposition for $(k+1)$-logarithmic integrals that
evaluate as Euler sums and hence explicitly in terms of Riemann zeta function
values at the positive integers.  Moreover, polygamma functions at the positive
integers are equivalent to the Hurwitz zeta function $\zeta(s,a)$ at the positive 
integers for both arguments.  

Our work can be extended to relations with the Lerch zeta function $\Phi(z,s,a)$.  
However, we restrict attention here to relations of much more immediate applications interest.  

We let $\zeta$ be the Riemann zeta function, Li$_s(z)$ the polylogarithm function
\cite{lewin}, and $H_n \equiv \sum_{k=1}^n 1/k$ the $n$th harmonic number.  Material
on special functions may be found in \cite{nbs,andrews,grad,coffeyjmp2}, and
especially on the dilogarithm Li$_2$ and polylogarithm functions in Lewin \cite{lewin58,lewin}.  

Our first result evaluates $(k+1)$-fold logarithmic integrals in terms of sums of
$\zeta$ values of weight $k+2$.  These integrals are equivalent to a non alternating
sum of polygamma function values.  We then evaluate $(j+1)$-fold logarithms integrals
that correspond to an alternating sum of polygamma function values, Proposition 2
giving the details of the $j=2$ case.  As a component of Proposition 4 for
the integral of a polylogarithm function and a rational factor, we explicitly obtain 
alternating parameterized polygamma function summations in two different ways.  
Proposition 5 subsumes Proposition 1 in that all integer powers of the logarithmic
factors are allowed.  In Proposition 6, we obtain parameterized polygamma sums and 
their equivalent multi-fold logarithmic integrals with rational factor.

\centerline{\bf Definite logarithmic integrals from Euler sums}

The definite integrals (44) of Coffey \cite{coffey05} are special cases of the following.

{\bf Proposition 1}.  For integers $k \geq 1$ we have
$$\int_0^1 {{\ln(1-t)} \over {1-t}}\ln^k t ~dt={{(-1)^{k+1}} \over 2}k!\left[(k+1)\zeta
(k+2)-\sum_{n=1}^{k-1} \zeta(k-n+1)\zeta(n+1)\right].  \eqno(1)$$

{\it Proof}.  Integrating by parts, we have
$$\int_0^1 {{\ln(1-t)} \over {1-t}}\ln^k t ~dt={k \over 2} \int_0^1 {{\ln^{k-1} t} \over t}
\ln^2(1-t) ~dt$$
$$=(-1)^{k+1}k! \sum_{r=1}^\infty {H_r \over {(r+1)^{k+1}}}.  \eqno(2)$$
Here we have used the standard generating function
$${1 \over 2}\ln^2(1-x)=\sum_{k=1}^\infty {H_k \over {k+1}} x^{k+1}, ~~~~|x|<1.  \eqno(3)$$
This equation immediately follows from integrating the standard generating function
for Li$_1(z)=-\ln(1-z)$.  In obtaining (2) we have also interchanged summation and
integration, justified on the basis of absolute convergence, and used the $\Gamma$
function integral
$$\int_0^1 t^r \ln^{k-1} t ~dt=(-1)^{k-1}\int_0^\infty u^{k-1}e^{-(r+1)u}du=(-1)^{k-1}
{{(k-1)!} \over {(r+1)^k}}.  \eqno(4)$$
We now use the functional equation of harmonic numbers,
$H_r=H_{r-1}+1/r$, to determine
$$\sum_{r=1}^\infty {H_r \over {(r+1)^{k+1}}}=\sum_{r=1}^\infty {H_r \over {r^{k+1}}}-\zeta(k+2).  \eqno(5)$$
We then apply Euler's relation for integers $m \geq 2$  
$$2\sum_{n=1}^\infty {H_n \over n^m}=(m+2)\zeta(m+1)-\sum_{n=1}^{m-2} \zeta(m-n)\zeta(n+1),
\eqno(6)$$
at $m=k+1$, and the Proposition follows.

{\it Further example}.  We have
$$\int_0^1 {{\ln(1-t)} \over {1-t}}\ln^5 t ~dt=-{4 \over 3}\pi^4 \zeta(3)-20\pi^2\zeta(5)
+360 \zeta(7).  \eqno(7)$$

{\it Remarks}.  There are several other representations for the Euler sums on the right
side of (2).  By using the inverse Laplace transform of $(r+1)^{-(k+1)}$ [cf. (4)] and 
interchanging summation and integration we have for $k \geq 1$
$$(-1)^{k+1}k! \sum_{r=1}^\infty {H_r \over {(r+1)^{k+1}}}=(-1)^k \int_0^\infty s^k {{\ln(1-e^{-s})} \over {e^s-1}}ds.  \eqno(8)$$

In the Proposition, we have an example of integrating an integrand of logarithmic 
weight $k+1$ that evaluates in terms of zeta values of weight $k+2$.

Relations with regard to sums over polygamma functions $\psi^{(j)}$ we largely omit.  However, we may readily verify that for $k \geq 1$,
$$(-1)^{k+1}k!\sum_{r=1}^\infty {H_r \over r^{k+1}}=\sum_{\ell=1}^\infty {{\psi^{(k)}(\ell)}
\over \ell}.  \eqno(9)$$
Upon interchanging sums, we have
$$\sum_{r=1}^\infty {H_r \over r^{k+1}}=\sum_{r=1}^\infty {1 \over r^{k+1}}\sum_{\ell=1}^r
{1 \over \ell}$$
$$=\sum_{\ell=1}^\infty {1 \over \ell}\sum_{r=\ell}^\infty {1 \over r^{k+1}}
=\sum_{\ell=1}^\infty {1 \over \ell}\sum_{r=0}^\infty {1 \over {(r+\ell)^{k+1}}}$$
$$=\sum_{\ell=1}^\infty {1 \over \ell}\zeta(k+1,\ell)={{(-1)^{k+1}} \over {k!}}
\sum_{\ell=1}^\infty {{\psi^{(k)}(\ell)} \over \ell}.  \eqno(10)$$

\medskip
\centerline{\bf Other integral expressions}

We first give expressions for the integral
$$I_2 \equiv \int_0^1 \ln(1+t) {{\ln^2 t} \over {1-t}}dt \simeq 0.345690556478. \eqno(11)$$
We have
$$I_2=\sum_{n=1}^\infty {{(-1)^n} \over n} \psi''(n+1)$$
$$=\sum_{n=1}^\infty {{(-1)^n} \over n} \psi''(n) -{7 \over{360}}\pi^4.  \eqno(12)$$
In obtaining the second line here we used the functional equation of the tetragamma
function, $\psi''(x+1)=\psi''(x)+2/x^3$, along with the alternating form of the
Riemann zeta function.  

Furthermore, we have
{\newline \bf Proposition 2}
$$I_2=\sum_{k=1}^\infty (-1)^k \int_0^\infty e^{-kt} \psi''(k)dt-{7 \over{360}}\pi^4$$
$$={7 \over 2}\ln 2 ~\zeta(3)-{{19} \over {720}}\pi^4.  \eqno(13)$$

{\it Proof}.  Here, with Li$_3$ the trilogarithm function, we have used the integral
$$\int_0^\infty {{\mbox{Li}_3(-e^{-t})} \over {1+e^t}}dt=\int_0^1 {{\mbox{Li}_3(-u)} \over
{1+u}}du={\pi^4 \over {288}}-{3 \over 4} \ln 2 ~\zeta(3).  \eqno(14)$$
We also recall the following relations.  We have 
$$\psi''(x)=-2\sum_{n=0}^\infty {1 \over {(n+x)^3}}.  \eqno(15)$$
Then we have
$$\sum_{k=1}^\infty (-1)^k e^{-kt}\psi''(k)=-2\sum_{k=1}^\infty (-1)^k e^{-kt}\sum_{n=0}^
\infty {1 \over {(n+k)^3}}$$
$$=-2\sum_{k=1}^\infty (-1)^k e^{-kt}\sum_{n=k}^\infty {1 \over n^3}
=-2\sum_{n=1}^\infty {1 \over n^3}\sum_{k=1}^n (-1)^k e^{-k t}$$
$$={{-2} \over {1+e^t}} \sum_{n=1}^\infty {1 \over n^3}[(-1)^n e^{-nt}-1]$$
$$={{-2} \over {1+e^t}} [\mbox{Li}_3(-e^{-t})-\zeta(3)].  \eqno(16)$$
For the integral (14) we employ integration by parts twice:
$$\int_0^1 \mbox{Li}_3(-u) \left({d \over {du}}\ln(1+u)\right)du
=-\int_0^1 {{\mbox{Li}_2(-u)} \over u}\ln(1+u)du-{3 \over 4}\ln 2 ~\zeta(3)$$
$$=-\int_0^1 \mbox{Li}_2(-u) \left({d \over {du}}\mbox{Li}_2(-u) \right)du-{3 \over 4}
\ln 2 ~\zeta(3)$$
$$=-{1 \over 2}\mbox{Li}_2^2(-u)|_0^1 -{3 \over 4}\ln 2 ~\zeta(3)$$
$$=-{1 \over 2}\left(-{\pi^2 \over {12}}\right)^2-{3 \over 4}\ln 2 ~\zeta(3).  \eqno(17)$$

{\it Remarks}.  Alternatively, we may employ the generating function
$$\sum_{n=1}^\infty (-1)^n H_n z^n=-{{\ln(1+z)} \over {1+z}}.  \eqno(18)$$
Then we obtain
$$I_2=-2\sum_{n=1}^\infty (-1)^n H_n\left[{1 \over {(n+1)^3}}+\psi''(n+1)\right]$$
$$=-2\sum_{n=1}^\infty (-1)^n H_n\left[{1 \over {(n+1)^3}}+{2 \over n^3}+\psi''(n)\right].
\eqno(19)$$

The integral $I_2$ appears in a semi-classical treatment of neutral atoms and diatomic molecules \cite{bernie}.  This theory employs improved Thomas-Fermi models
and provides chemically accurate results.  It enables a better understanding of the variation of properties within, for instance, a series of homologous compounds by
including the number of electrons as a variable.   
It is within this kind of framework that various sums and integrals have arisen and require closed forms for the theory's most exact expression.

We have the following Corollary from Proposition 2.
{\newline \bf Corollary 1}.  A closed form for the integral $\int_0^1 \ln(1-t){{\ln^2 t}
\over {1+t}}dt$ determines one for the integral $\int_0^1 {{\ln t} \over t}\ln(1-t) \ln(1+t)dt$, and vice versa.

{\it Proof}.  By integrating by parts, we have from the definition (11)
$$I_2=\int_0^1 \ln(1-t)\left[{{\ln^2 t} \over {1+t}}+2{{\ln t} \over t}\ln(1+t)\right]dt.
$$  
Since $I_2$ is known, the result follows.

Expressions for the integrals of Corollary 1 follow from the considerations given in the
Appendix.

\pagebreak
\centerline{\bf Generalization}
\medskip

It is evident that the method of Proposition 2 extends to a family
of integrals for the ``lower sign" of (43) of \cite{coffey05}.  We have
$$\psi^{(n)}(x)=(-1)^{n+1}n!\sum_{k=0}^\infty {1 \over {(x+k)^{n+1}}}.  \eqno(20)$$
Then we obtain
$$\sum_{k=1}^\infty (-1)^k e^{-kt} \psi^{(n)}(k)={{(-1)^{n+1}n!} \over {1+e^t}}
\left[\mbox{Li}_{n+1}(-e^{-t})-\zeta(n+1)\right].  \eqno(21)$$
We use this in the proof of
\newline{\bf Proposition 3}.  Put
$$\sum_{n=1}^\infty {{(-1)^n} \over n} \psi^{(j)}(n+1)=\int_0^1 \ln^j t {{\ln(1+t)} \over
{1-t}}dt \equiv I_j.  \eqno(22)$$
Then for integers $j\geq 1$ we have
$$I_j=(-1)^{j+1}j!\int_0^1 {{\mbox{Li}_{j+1}(-u)} \over {1+u}}du +(-1)^jj![\ln 2~\zeta(j+1)
+(2^{-j-1}-1)\zeta(j+2)].  \eqno(23)$$

{\it Proof}.  The equality of the summation and integration in (22) may be verified in
many ways, including simply expanding the factor $\ln(1+t)$ of the integrand in power
series:
$$I_j=\sum_{\ell=1}^\infty {{(-1)^{\ell+1}} \over \ell}\int_0^1 t^\ell {{\ln^j t} \over
{1-t}}dt$$
$$=\sum_{\ell=1}^\infty {{(-1)^{\ell+1}} \over \ell} (-1)^j j!\zeta(j+1,\ell+1)$$
$$=\sum_{\ell=1}^\infty {{(-1)^\ell} \over \ell} \psi^{(j)}(\ell+1).  \eqno(24)$$
The interchange of summation and integration is justified by absolute convergence. 

We have the functional equation
$$\psi^{(j)}(x+1)=\psi^{(j)}(x)+{{(-1)^j j!} \over x^{j+1}}.  \eqno(25)$$
Using this relation and proceeding as in the proof of the previous Proposition, we have
$$I_j=\sum_{n=1}^\infty (-1)^n \int_0^\infty e^{-nt} \psi^{(j)}(n)dt+(-1)^jj!\sum_{n=1}^
\infty {{(-1)^n} \over n^{j+2}}$$
$$=\sum_{n=1}^\infty (-1)^n \int_0^\infty e^{-nt} \psi^{(j)}(n)dt+(-1)^jj!(2^{-j-1}-1)
\zeta(j+2), \eqno(26)$$
where we used the alternating form of the Riemann zeta function, Li$_j(-1)=(2^{1-j}-1)
\zeta(j)$.  From relations (21) and (26) we obtain
$$(-1)^{j+1}j! \int_0^\infty {1 \over {1+e^t}} \left[\mbox{Li}_{j+1}(-e^{-t})-\zeta(j+1)
\right] dt+(-1)^jj!(2^{-j-1}-1)\zeta(j+2)$$
$$=(-1)^{j+1}j! \int_0^1 {{\left[\mbox{Li}_{j+1}(-u)-\zeta(j+1)\right]} \over {1+u}} du +(-1)^jj!(2^{-j-1}-1)\zeta(j+2), \eqno(27)$$
from which the Proposition follows.

{\it Remarks}.  Not surprisingly, the polylogarithmic integrals of Proposition 3 may 
be written in terms of sums over generalized harmonic numbers $H_n^{(r)} \equiv
\sum_{k=1}^n 1/k^r$.  We have the generating function
$$\sum_{n=1}^\infty (-1)^n H_n^{(r)} z^n={{\mbox{Li}_r(-z)} \over {1+z}}, ~~~~|z|<1.
\eqno(28)$$
Then using the functional equation $H_n^{(j+1)}=H_{n+1}^{(j+1)}-1/(n+1)^{j+1}$ we obtain
$$\int_0^1 {{\mbox{Li}_{j+1}(-u)} \over {1+u}}du=-\sum_{n=2}^\infty {{(-1)^n} \over n}
H_n^{(j+1)} +1+(2^{-j-1}-1)\zeta(j+2).  \eqno(29)$$


We next present an evaluation of the polylogarithmic integrals of Proposition 3.  We
have
{\newline \bf Proposition 4}.  For integers $j \geq 1$ we have (a)
$$\int_0^1 {{\mbox{Li}_{j+1}(-u)} \over {1+u}}du=(-1)^j \sum_{k=1}^\infty {{(-1)^{k-1}}
\over k^{j+1}}\Phi(-1,1,k+1)+[(-1)^j+1](2^{-j}-1)\ln 2 ~\zeta(j+1)$$
$$+(-1)^j\sum_{m=1}^{j-1}(-1)^m(2^{-m}-1)\zeta(m+1)(1-2^{m-j})\zeta(j+1-m).  \eqno(30)$$
(b) Of note, when $j$ is odd, we have
$$\int_0^1 {{\mbox{Li}_{j+1}(-u)} \over {1+u}}du=(-1)^j \sum_{k=1}^\infty {{(-1)^{k-1}}
\over k^{j+1}}\Phi(-1,1,k+1).  \eqno(31a)$$
When $j$ is even, we have
$$\int_0^1 {{\mbox{Li}_{j+1}(-u)} \over {1+u}}du=\sum_{k=1}^\infty {{(-1)^{k-1}}
\over k^{j+1}}\Phi(-1,1,k+1)+2(2^{-j}-1)\ln 2 ~\zeta(j+1)$$
$$+2\sum_{m=1}^{j/2-1}(-1)^m(2^{-m}-1)\zeta(m+1)(1-2^{m-j})\zeta(j+1-m)
+(-1)^{j/2-1}(1-2^{-j/2})^2\zeta^2(j/2+1).  \eqno(31b)$$
(c)  We have the alternative representation
$$\int_0^1 {{\mbox{Li}_{j+1}(-u)} \over {1+u}}du=\ln 2 ~\zeta(j+1)-{1 \over {j!}}\int_0^\infty {{t^j e^t} \over {e^t-1}}\ln(1+e^{-t})dt  \eqno(32)$$
$$=\ln 2 ~\zeta(j+1)+[2^{-(j+1)}-1]\zeta(j+2)+{{(-1)^{j-1}} \over {j!}}\int_0^1 \ln^j v
{{\ln(1+v)} \over {1-v}}dv.  \eqno(33)$$

{\it Proof}.  By expanding the integrand factor $(1+u)^{-1}$ as a geometric series,
by integrating term-by-term we have
$$\int_0^1 {{\mbox{Li}_{j+1}(-u)} \over {1+u}}du=\sum_{k=0}^\infty (-1)^k \int_0^1
u^k \mbox{Li}_{j+1}(-u)du$$
$$=\sum_{k=0}^\infty (-1)^k \sum_{n=1}^\infty {{(-1)^n} \over n^{j+1}}\int_0^1 u^{k+n}du$$
$$=\sum_{k=0}^\infty (-1)^k \sum_{n=1}^\infty {{(-1)^n} \over n^{j+1}}{1 \over {(k+n+1)}}$$
$$=\sum_{k=0}^\infty (-1)^k S_j(k), \eqno(34)$$
where we have defined for $z \in C/\{-2,-3,\ldots\}$,
$$S_j(z) \equiv \sum_{n=1}^\infty {{(-1)^n} \over n^{j+1}}{1 \over {(n+z+1)}}.  \eqno(35)$$

We next explicitly determine $S_j(z)$ by writing a recursion relation for it.
We have
$$S_j(z)={1 \over {(z+1)}}\sum_{n=1}^\infty {{(-1)^n} \over n^j}\left({1 \over n}-{1 \over
{n+z+1}}\right), \eqno(36)$$
and find
$$S_j(z)=-{{S_{j-1}(z)} \over {z+1}}+{1 \over {(z+1)}}(2^{-j}-1)\zeta(j+1), \eqno(37a)$$
with
$$S_0(z)={1 \over {(z+1)}}[\Phi(-1,1,z+2)-\ln 2].  \eqno(37b)$$
Iteration of the recurrence (37a) gives
$$S_j(z)=(-1)^j {{S_0(z)} \over {(z+1)^j}}+{{(-1)^j} \over {(z+1)^{j+1}}}\sum_{m=1}^j
(-1)^m (2^{-m}-1)(z+1)^m \zeta(m+1).  \eqno(38)$$
We next substitute (38) in (34) and perform sums with the alternating form of the
Riemann zeta function,
$$\sum_{k=0}^\infty (-1)^k S_j(k)=(-1)^j\left[\sum_{k=1}^\infty{{(-1)^{k-1}} \over k^{j+1}}
\Phi(-1,1,k+1)-\ln 2(1-2^{-j})\zeta(j+1)\right]$$
$$+(-1)^j\sum_{m=1}^j (-1)^m (2^{-m}-1)\zeta(m+1)(1-2^{m-j})\zeta(j+1-m).  \eqno(39)$$
We apply the property 
$$\lim_{x\to 0}(1-2^x)\zeta(1-x)=\ln 2, \eqno(40)$$ 
for the $m=j$ term in the second line of (39).  Then combining $(2^{-j}-1)\ln 2 ~\zeta(j+1)$ terms gives the the final form of part (a) of the Proposition.

For part (c) we employ the integral representation
$$\mbox{Li}_s(z)={z \over {\Gamma(s)}}\int_0^\infty {t^{s-1} \over {e^t-z}}dt,  \eqno(41)$$
and interchange integrations.  By using partial fractions we have
$$-\int_0^1 {{u du} \over {(u+1)(u+e^t)}}={1 \over {1-e^t}}\int_0^1\left({e^t \over {u+
e^t}}-{1 \over {u+1}}\right)du={1 \over {e^t-1}}[\ln 2+te^t-e^t \ln(1+e^t)].  \eqno(42)$$
Then using (41) for the special case of $z=1$ gives (32).  With the change of variable
$t=-\ln v$, the use of partial fractions, 
$$\int_0^1 {{\mbox{Li}_{j+1}(-u)} \over {1+u}}du=\ln 2 ~\zeta(j+1)+{{(-1)^j} \over {j!}}
\int_0^1\left({1 \over v}+{1 \over {1-v}}\right)\ln^j v \ln(1+v)dv, \eqno(43)$$
and lastly integration by parts we obtain (33).

{\it Remarks}.  Given the definition (22) of $I_j$, it is seen that (33) is equivalent
to (23).  

We may note the limit relation
$$\lim_{s \to \infty}\int_0^1 {{\mbox{Li}_{s+1}(-u)} \over {1+u}}du=-\int_0^1 {u \over
{1+u}}du=\ln 2-1.  \eqno(44)$$

The value $\Phi(-1,1,2)=1-\ln 2$ and (35)-(36) reduce properly to the Proposition
\cite{coffey05} (p. 96-97; (59), (61)) at $z=0$.  We have
$$\lim_{j \to \infty}S_j(z)=-{1 \over {z+2}}.  \eqno(45)$$

Another method to determine the sums $S_j(z)$ proceeds from the relation
$$\sum_{n=1}^\infty {{(-1)^n} \over {(n+b)}}{1 \over {(n+z)}}={1 \over {(b-z)}}
[\Phi(-1,1,b+1)-\Phi(-1,1,z+1)].  \eqno(46)$$
This equation follows simply from partial fractions and the series definition of
the Lerch zeta function.  The terms on the right side of (41) are nothing but a form
of the alternating Hurwitz zeta function, as
$$\sum_{n=1}^\infty {{(-1)^n} \over {n+b}}=-\Phi(-1,1,b+1)={1 \over 2}\left[\psi\left(
{{b+1} \over 2}\right)-\psi\left({b \over 2}+1\right)\right].  \eqno(47)$$
We have
$${\partial \over {\partial b}}\Phi(-1,1,b+1)={1 \over 4}\left[\zeta\left(2,{b \over 2}
+1\right)-\zeta\left(2,{{b+1} \over 2}\right)\right], \eqno(48)$$
and $\partial_a \zeta(s,a)=-s\zeta(s+1,a)$, giving
$${\partial^j \over {\partial b^j}}\Phi(-1,1,b+1)=(-1)^{j+1}{{j!} \over 2^{j+1}}\left[ \zeta\left(j+1,{b \over 2}+1\right)-\zeta\left(j+1,{{b+1} \over 2}\right)\right]. \eqno(49)$$
We then have
$${\partial^j \over {\partial b^j}}\sum_{n=1}^\infty {{(-1)^n} \over {(n+b)}}{1 \over {(n+z)}}=(-1)^j j!\sum_{n=1}^\infty {{(-1)^n} \over {(n+b)^{j+1}}}{1 \over {(n+z)}}$$
$$={\partial^j \over {\partial b^j}}{1 \over {(b-z)}}[\Phi(-1,1,b+1)-\Phi(-1,1,z+1)]$$
$$=\sum_{m=0}^j {j \choose m}\left[\left({\partial \over {\partial b}}\right)^{j-m}
{1 \over {(b-z)}}\right]\left({\partial \over {\partial b}}\right)^m \Phi(-1,1,b+1)
-(-1)^j j! {{\Phi(-1,1,b+1)} \over {(b-z)^{j+1}}}$$
$$=\sum_{m=0}^j {j \choose m}{{(j-m)!(-1)^{j-m}} \over {(b-z)^{j-m+1}}}{{(-1)^{m+1}
m!} \over 2^{m+1}}\left[ \zeta\left(m+1,{b \over 2}+1\right)-\zeta\left(m+1,{{b+1} \over 2}\right)\right]$$
$$-(-1)^j j! {{\Phi(-1,1,b+1)} \over {(b-z)^{j+1}}}$$
$$=j!\sum_{m=0}^j {{(-1)^{j-m}} \over {(b-z)^{j-m+1}}}{{(-1)^{m+1}} \over 2^{m+1}}\left[ \zeta\left(m+1,{b \over 2}+1\right)-\zeta\left(m+1,{{b+1} \over 2}\right)\right]$$
$$-(-1)^j j! {{\Phi(-1,1,b+1)} \over {(b-z)^{j+1}}}.  \eqno(50)$$

We now put $b=0$ and cancel factors $(-1)^jj!$ on both sides of (50), so that
$$S_j(z-1)=\sum_{n=1}^\infty {{(-1)^n} \over n^{j+1}}{1 \over {(n+z)}}=\sum_{m=0}^j {{(-1)^{j-m}} \over {(-z)^{j-m+1}}}{{(-1)^{m+1}} \over 2^{m+1}}\left[ \zeta\left(m+1,{b \over 2}+1\right)-\zeta\left(m+1,{{b+1} \over 2}\right)\right]$$
$$-{{\Phi(-1,1,b+1)} \over {(-z)^{j+1}}}.  \eqno(51)$$
By using the relation $\zeta(s,1/2)=(2^s-1)\zeta(s)$, we have
$$S_j(z-1)={{(-1)^j} \over z^{j+1}}\sum_{m=0}^j (-1)^m z^m(2^{-m}-1)\zeta(m+1)+{{(-1)^j}
\over z^{j+1}}\Phi(-1,1,z+1)$$
$$={{(-1)^j} \over z^{j+1}}\sum_{m=1}^j (-1)^m z^m(2^{-m}-1)\zeta(m+1)+{{(-1)^{j+1}}
\over z^{j+1}}\ln 2 + {{(-1)^j} \over z^{j+1}}\Phi(-1,1,z+1), \eqno(52)$$
where we again applied the property (40) for the $m=0$ term.  We recover the
expression (38) for $S_j(z)$.

We may generalize the sum of (35) to 
$$S_j(t,z) \equiv \sum_{n=1}^\infty {t^n \over n^{j+1}}{1 \over {(n+z+1)}}, ~~~~|t|
\leq 1,  \eqno(53)$$
with
$$S_0(t,z)={1 \over {z+1}}[-t\Phi(t,1,z+2)-\ln(1-t)].  \eqno(54)$$

We may explicitly determine $S_j(t,z)$.  We do so, and record this as the following
Lemma, as this result in the special case of $t=-1$ could have been used in the proof of Proposition 4.   We have
{\newline \bf Lemma 1}.  For integers $j \geq 1$ we have
$$S_j(t,z)={{(-1)^j} \over {(z+1)^j}}S_0(t,z)+{{(-1)^j} \over {(z+1)^{j+1}}}\sum_{m=1}^j
(-1)^m (z+1)^m \mbox{Li}_{m+1}(t).  \eqno(55)$$

{\it Proof}.  We have
$$S_j(t,z)={1 \over {(z+1)}}\sum_{n=1}^\infty {t^n \over n^j}\left({1 \over n}-{1 \over
{n+z+1}}\right), \eqno(56)$$
and find the recurrence
$$S_j(t,z)=-{{S_{j-1}(t,z)} \over {z+1}}+{1 \over {z+1}}\mbox{Li}_{j+1}(t).  \eqno(57)$$
Iterating this relation yields the Lemma.

Furthermore, we have the following extension subsuming Lemma 1.  We have
{\newline \bf Lemma 2}.  Put
$$S_j(t,z,a) \equiv \sum_{n=1}^\infty {t^n \over {(n+a)^{j+1}}}{1 \over {(n+z+1)}}, ~~~~|t|
\leq 1,  \eqno(58)$$
with
$$S_0(t,z,a)={t \over {z-a+1}}[\Phi(t,1,a+1)-\Phi(t,1,z+2)].  \eqno(59)$$
Then we have for integers $j \geq 1$
$$S_j(t,z,a)={{(-1)^j} \over {(z-a+1)^j}}S_0(t,z,a)+{{(-1)^j t} \over {(z-a+1)^{j+1}}}\sum_{m=1}^j (-1)^m (z-a+1)^m \Phi(t,m+1,a+1).  \eqno(60)$$

{\it Proof}.  The proof is very similar to that for the preceding Lemma and is omitted.

We next show that a generalization of Proposition 1 to all integer powers of the
logarithmic factors can be evaluated in terms of zeta values at the integers.  We have
{\newline \bf Proposition 5}.  Put the function
$$z(x,y) \equiv \sum_{k,j=1}^\infty (-1)^{k+j+1} {{(k+j-1)!} \over {k!j!}}\zeta(k+j)
x^k y^j.  \eqno(61)$$
Then we have
$$\int_0^1 {{\ln^k t} \over t}\ln^j(1-t) ~dt=\left.\left({\partial \over {\partial x}}\right)^k \left({\partial \over {\partial y}}\right)^j \right|_{x=y=0} {1 \over x}
\exp(z).  \eqno(62)$$

{\it Proof}.  We make use of the Beta function integral for Re $x>0$ and Re $y>-1$,
$$\int_0^1 t^{x-1}(1-t)^ydt=B(x,y+1)={{\Gamma(x)\Gamma(y+1)} \over {\Gamma(x+y+1)}}$$
$$={{\Gamma(x+1)\Gamma(y+1)} \over {x\Gamma(x+y+1)}}.  \eqno(63)$$
Then we have
$$\int_0^1 {{\ln^k t} \over t}\ln^j(1-t) ~dt=\left.\left({\partial \over {\partial x}}\right)^k \left({\partial \over {\partial y}}\right)^j \right|_{x=y=0}={1 \over x}
e^z, \eqno(64)$$
where we put
$z(x,y)=\ln\Gamma(x+1)+\ln \Gamma(y+1)-\ln \Gamma(x+y+1)$.  For $k\geq 1$ and $j \geq 1$
we have
$$\partial_x^k \partial_y^j z =-\psi^{(k+j-1)}(x+y+1)$$
$$~~~~~~~~~~~~~~~~~~~=(-1)^{k+j+1} (k+j-1)!\zeta(k+j,x+y+1), \eqno(65)$$
where we used relation (20).  Noting that $z(0,0)=z(x,0)=z(0,y)=0$, we have the
Maclaurin series
$$z(x,y)=\sum_{k=1}^\infty \sum_{j=1}^\infty {{\partial_x^k \partial_y^j|_{x=y=0}z}
\over {k! j!}} x^k y^j=\sum_{k=1}^\infty \sum_{j=1}^\infty (-1)^{k+j+1}{{(k+j-1)!}\over {k! j!}} \zeta(k+j) x^k y^j.  \eqno(66)$$
The Proposition follows.

{\it Remark}.  In the Proposition we have developed a $2$-variable generating function.
A generalization for the integrals
$$\int_0^1 {{\ln^j t} \over t}\ln^k(1-t) \ln^\ell(1-\alpha t)~dt \equiv I_{jk\ell}$$
is for Re $u>0$, Re $v>-1$,
$$\sum_{j,k,\ell=0}^\infty {{u^j v^k (-w)^\ell} \over {j! k! \ell!}} \int_0^1 {{\ln^j t} \over t}\ln^k(1-t) \ln^\ell(1-\alpha t)~dt=\int_0^1 t^{u-1}(1-t)^v (1-\alpha t)^{-w}dt$$
$$=B(u,v+1) ~_2F_1(u,w;1+u+v;\alpha),  \eqno(67)$$
where $_2F_1$ is the Gauss hypergeometric function.  I.e., we have
$$I_{jk\ell}=\left.\left({\partial \over {\partial u}}\right)^j\left({\partial \over {\partial v}}\right)^k\left(-{\partial \over {\partial w}}\right)^\ell \right|_{u=v=w=0}
B(u,v+1) ~_2F_1(u,w;1+u+v;\alpha).  \eqno(68)$$

Another generalization of Proposition 1 is to consider logarithmic integrals with sign alteration
of the integrand.  For this we let $_kF_\ell$ be the generalized hypergeometric function (e.g.,
\cite{andrews}).  We have
{\newline \bf Proposition 6}.  Let $|z| \leq 1$ and $p \geq 1$ be an integer.  Then we have (a)
$$\sum_{k=1}^\infty {{(-1)^k} \over k^p} \psi(k)z^k=-\gamma \mbox{Li}_p(-z)+z \left.{\partial \over
{\partial \beta}}\right|_{\beta=1} ~_{p+2}F_{p+1}(1,\ldots,1;2,\ldots,2,\beta;-z),  \eqno(69)$$
and (b) 
$$\sum_{k=1}^\infty {{(-1)^k} \over k^p} \psi(k)=-\gamma (1-2^{1-p})\zeta(p)+\left.{\partial \over
{\partial \beta}}\right|_{\beta=1} ~_{p+2}F_{p+1}(1,\ldots,1;2,\ldots,2,\beta;-1).  \eqno(70)$$

{\it Proof}.  Our proof provides an alternative integral representation.  We have
$$\sum_{k=1}^\infty {{(-1)^k} \over k^p} \psi(k)z^k={1 \over {\Gamma(p)}} \sum_{k=1}^\infty (-1)^k \psi(k) z^k \int_0^\infty t^{p-1} e^{-kt} dt$$
$$={z \over {\Gamma(p)}} \int_0^\infty t^{p-1} {{[\gamma+\ln(1+ze^{-t})]} \over {z+e^t}}dt$$
$$={z \over {\Gamma(p)}}\int_0^1 \ln^{p-1}\left({1 \over v}\right) {{[\gamma+\ln(1+zv)]} \over
{1+zv}}dv$$
$$=-\gamma \mbox{Li}_p(-z) -{z \over {\Gamma(p)}}\sum_{n=1}^\infty (-1)^n H_n z^n \int_0^1 v^n
\ln^{p-1}\left({1 \over v}\right)dv, \eqno(71)$$
where we have used (18).  Carrying out the integral gives
$$\sum_{k=1}^\infty {{(-1)^k} \over k^p} \psi(k)z^k=-\gamma \mbox{Li}_p(-z) -z\sum_{n=1}^\infty
{{(-1)^n H_n z^n} \over {(n+1)^p}}.  \eqno(72)$$
We now consider the function
$$~_{p+2}F_{p+1}(1,\ldots,1;2,\ldots,2,\beta;-z)=\sum_{j=0}^\infty {{(1)_j^{p+2}} \over {(2)_j^p}}
{1 \over {(\beta)_j}} {{(-z)^j} \over {j!}}=\sum_{j=0}^\infty {1 \over {(j+1)^p}} {{j!} \over 
{(\beta)_j}}(-z)^j, \eqno(73)$$
where the Pochhammer symbol $(w)_n=\Gamma(w+n)/\Gamma(w)$, giving
$${\partial \over {\partial \beta}}~_{p+2}F_{p+1}(1,\ldots,1;2,\ldots,2,\beta;-z)=\sum_{j=1}^\infty 
{1 \over {(j+1)^p}} {{j!} \over {(\beta)_j}}(-z)^j[\psi(\beta)-\psi(\beta+j)].  \eqno(74)$$
Therefore, we find
$$\left.{\partial \over {\partial \beta}}\right|_{\beta=1} ~_{p+2}F_{p+1}(1,\ldots,1;2,\ldots,2,\beta;-z)
=-\sum_{j=1}^\infty {{(-z)^j} \over {(j+1)^p}} [\gamma+\psi(j+1)]=-\sum_{j=1}^\infty {H_j \over
{(j+1)^p}} (-z)^j.  \eqno(75)$$
Combining (72) and (75) yields part (a) of the Proposition.  Part (b) follows at $z=1$ by using
the relation Li$_p(-1)=-(1-2^{1-p})\zeta(p)$.

{\it Remarks}.  When $z=1$ and $p=1$ in Proposition 5, the term $\gamma (1-2^{1-p})\zeta(p)$ gives $-\gamma \ln 2$.  

The special case of $z=1$ and $p=5$ occurs in quantum chemistry calculations for
multi-electron atoms \cite{bernie}.

Equation (72) may be directly obtained from the defining sum on the left side of (69), but this route omits the integral representation that may be of interest in its own right.

Higher order derivatives of $_kF_\ell$ may be used to evaluate sums with generalized harmonic numbers.  However, this leads to nonlinear sums.  As an example, two derivatives of the generalized hypergeometric function of Proposition 6 generates not just $H_j^{(2)}$ but $H_j^2$.

An extension of Proposition 6 is to consider sums of the form for Re $a \geq 0$
$$\sum_{k=1}^\infty {{(-1)^k} \over {(k+a)^p}} \psi(k)z^k={{\gamma z} \over {\Gamma(p)}}\int_0^1 v^a {{\ln^{p-1}(1/v)} \over {1+zv}}dv-z \sum_{n=1}^\infty 
{{(-1)^n H_n z^n} \over {(n+a+1)^p}}.  \eqno(76)$$

\medskip
\centerline{\bf Appendix:  Other triple-logarithm integrals}
\medskip

Here we consider the integrals of Corollary 1.  We let $\gamma=-\psi(1)$ be the
Euler constant, and recall that the integral $I_2$ is defined in (11).  We first have
{\newline \bf Proposition A1}.  We have
$$\int_0^1 \ln(1-x){{\ln^2 x}\over {1+x}}dx=-{{7\pi^4} \over {180}}-{3 \over 2}\gamma
\zeta(3)+2\sum_{j=1}^\infty{{(-1)^j} \over j^3}[\psi(j)-j\psi'(j)]+I_2.  \eqno(A.1)$$
 
{\it Proof}.  We first note that the integral
$$\int_0^1 x^a \ln(1-x)\ln^2 x ~dx=-{1 \over {(a+1)^4}}[4+2(a+1)\gamma+2(a+1)\psi(a+1)$$
$$-2(a+1)^2 \psi'(a+1)+(a+1)^3 \psi''(a+2)], ~~~~~~\mbox{Re} ~a>-1, \eqno(A.2)$$
may be obtained from logarithmic differentiation twice with respect $a$ of the integral
$$\int_0^1 x^a \ln(1-x)dx=-{1 \over {a+1}}H_{a+1}=-{1 \over {a+1}}[\psi(a+2)+\gamma],
~~~~~~ \mbox{Re} ~a>-1, \eqno(A.3)$$
along with the functional equations of $\psi$ and $\psi'$.  We then expand the
integral of the left side of (A.1) as
$$\int_0^1 \ln(1-x){{\ln^2 x}\over {1+x}}dx=\sum_{j=0}^\infty (-1)^j \int_0^1 x^j
\ln(1-x)\ln^2 x ~dx$$
$$=-\sum_{j=0}^\infty {{(-1)^j} \over {(j+1)^4}}[4+2(j+1)\gamma+2(j+1)\psi(j+1)$$
$$-2(j+1)^2 \psi'(j+1)+(j+1)^3 \psi''(j+2)]. \eqno(A.4)$$
By using the summation form of $I_2$ in (12) and the alternating form of the Riemann
zeta function, the Proposition follows.

For the polygammic summations in (A.1), we have the following integral representations:
$$\sum_{j=1}^\infty {{(-1)^j} \over j^3}\psi(j+1)={1 \over 4}\int_0^1 [4\mbox{Li}_3(-t)+3\zeta(3)]{{dt} \over {t-1}}+{3 \over 4} \gamma \zeta(3),  \eqno(A.5)$$
and
$$\sum_{j=1}^\infty {{(-1)^j} \over j^2}\psi'(j)=\int_0^1 {{\mbox{Li}_2(-t)} \over
{t(t-1)}}\ln t ~dt.  \eqno(A.6)$$
These follow from using a standard integral representation for $\psi$ \cite{grad} (p. 943).
They may be rewritten in a number of ways by using integration by parts.

In addition, we have
$$\sum_{j=1}^\infty {{(-1)^j} \over j^3}\psi(j)={1 \over 2}\sum_{j=1}^\infty (-1)^j
\psi(j) \int_0^\infty t^2 e^{-jt}dt$$
$$={1 \over 2} \int_0^\infty {t^2 \over {1+e^t}}[\gamma+\ln(1+e^{-t})]dt
={1 \over 2} \int_0^1 {{\ln^2 u} \over {1+u}}[\gamma+\ln(1+u)]du$$
$$={3 \over 4}\gamma \zeta(3)+{1 \over {48}}\left[-\pi^4 -4\pi^2 \ln^2 2+4 \ln^4 2+96
\mbox{Li}_4\left({1 \over 2}\right)+84\ln 2 ~\zeta(3)\right].  \eqno(A.7)$$
This result may also be obtained from (30) of \cite{coffey05}.
Moreover,
$$\sum_{j=1}^\infty {{(-1)^j} \over j^2}\psi'(j)=\sum_{j=1}^\infty (-1)^j
\psi'(j) \int_0^\infty t e^{-jt}dt$$
$$=-{5 \over 4}\zeta(4)+\int_0^\infty {{t\mbox{Li}_2(-e^{-t})} \over {1+e^t}}dt$$
$$=-{5 \over 4}\zeta(4)-\int_0^1 {{\mbox{Li}_2(-u)\ln u} \over {1+u}}du, \eqno(A.8)$$
whereby integrating by parts we have
$$\int_0^1 {{\mbox{Li}_2(-u)\ln u} \over {1+u}}du=-\int_0^1 [\ln(1+u)\ln u-\mbox{Li}_2(-u)] {{\ln(1+u)} \over u}du$$
$$={1 \over {24}}\left[\pi^4 +4\pi^2 \ln^2 2-4 \ln^4 2-96\mbox{Li}_4 \left({1 \over 2}\right)-84\ln 2 ~\zeta(3)\right] +{\pi^4 \over {288}}.  \eqno(A.9)$$

By expanding the factor $\ln(1-x)$ of the left side of (A.1) in power series we also
have
$$\int_0^1 \ln(1-x){{\ln^2 x}\over {1+x}}dx=-{1 \over 8}\sum_{k=1}^\infty {1 \over k}
\left[\psi''\left({k \over 2}+1\right)-\psi''\left({{k+1} \over 2}\right)\right].
\eqno(A.10)$$
Related tetragamma series are given by
$$\sum_{k=1}^\infty {1 \over k}\psi''\left({k \over 2}+1\right)=\sum_{k=1}^\infty {1 \over k}\psi''\left({k \over 2}\right)+16 \zeta(4), \eqno(A.11)$$
and
$${1 \over 8}\sum_{k=1}^\infty {1 \over k}\left[\psi''\left({k \over 2}\right) +\psi''\left({{k+1} \over 2}\right)\right]=\sum_{k=1}^\infty {{\psi''(k)} \over k}=
-{5 \over 2}\zeta(4).  \eqno(A.12)$$
For the latter relation, we have applied the duplication formula
$$\psi''(2x)={1 \over 8}\left[\psi''(x)+\psi''\left(x+{1 \over 2}\right)\right].  \eqno(A.13)$$

We may use (42) to show the equivalence of the right side of (A.7) to Proposition A1.
For by applying the operator $\partial_b^2$ to both sides of (42) we have
$$2\sum_{n=1}^\infty {{(-1)^n} \over {(n+b)^3}}={1 \over 8}\left[\psi''\left(
{{b+1} \over 2}\right)-\psi''\left({b \over 2}+1\right)\right].  \eqno(A.14)$$
We then have with the aid of partial fractions
$${1 \over 8}\sum_{b=1}^\infty {1 \over b}\left[\psi''\left({{b+1} \over 2}\right) -\psi''\left({b \over 2}+1\right)\right]=2\sum_{b=1}^\infty {1 \over b}\sum_{n=1}^\infty {{(-1)^n} \over {(n+b)^3}}$$
$$=\sum_{n=1}^\infty {{(-1)^n} \over {n^3}}[2\gamma+2\psi(n+1)-2n\psi'(n+1)+n^2\psi''(n+1)]
$$
$$=-{3 \over 2}\gamma \zeta(3)+2\sum_{n=1}^\infty {{(-1)^n} \over {n^3}}[\psi(n+1)-n
\psi'(n+1)]+I_2$$
$$=-{3 \over 2}\gamma \zeta(3)+2\sum_{n=1}^\infty {{(-1)^n} \over {n^3}}[\psi(n)-n
\psi'(n)] +4 \sum_{n=1}^\infty{{(-1)^n} \over {n^4}}+I_2$$
$$=-{{7\pi^4} \over {180}}-{3 \over 2}\gamma \zeta(3)+2\sum_{n=1}^\infty{{(-1)^n} \over n^3}[\psi(n)-n\psi'(n)]+I_2.  \eqno(A.15)$$

\pagebreak

\end{document}